# Engineering of Niobium Surfaces Through Accelerated Neutral Atom Beam Technology For Quantum Applications

Soumen Kar[1,*], Conan Weiland[2], Chenyu Zhou[3], Ekta Bhatia[1], Brian Martinick[1], Jakub Nalaskowski[1], John Mucci[1], Stephen Olson[1], Pui Yee Hung[1], Ilyssa Wells[1], Hunter Frost[4], Corbet S. Johnson[1], Thomas Murray[1], Vidya Kaushik[1], Sean Kirkpatrick[5], Kiet Chau[5], Michael J. Walsh[5], Mingzhao Liu[3], Satyavolu S. Papa Rao[1,4,#]

[1]NY CREATES, Albany, NY 12203

[2]Material Measurement Laboratory, National Institute of Standards and Technology, Gaithersburg, MD 20899

[3]Center for Functional Nanomaterials, Brookhaven National Laboratory, Upton, NY 19973

[4]College of Nanoscale Science and Engineering, SUNY Polytechnic Institute, Albany, NY 12203

[5]Neutral Physics Corporation, Billerica, MA 01821

*Email:* *skar@sunypoly.edu, #spaparao@sunypoly.edu

*Abstract* – A major roadblock to scalable quantum computing is phase decoherence and energy relaxation caused by qubits interacting with defect-related two-level systems (TLS). Native oxides present on the surfaces of superconducting metals used in quantum devices are acknowledged to be a source of TLS that decrease qubit coherence times. Reducing microwave loss by "surface engineering" (*i.e.,* replacing uncontrolled native oxide of superconducting metals with a thin, stable surface with predictable characteristics) can be a key enabler for pushing performance forward with devices of higher quality factor. In this work, we present a novel approach to replace the native oxide of niobium (typically formed in an uncontrolled fashion when its pristine surface is exposed to air) with an engineered oxide, using a room-temperature process that leverages Accelerated Neutral Atom Beam (ANAB) technology at 300 mm wafer scale. This ANAB beam is composed of a mixture of argon and oxygen, with tunable energy per atom, which is rastered across the wafer surface. The ANAB-engineered Nb-oxide thickness was found to vary from 2 nm to 6 nm depending on ANAB process parameters. Modeling of variable-energy XPS data confirm thickness and compositional control of the Nb surface oxide by the ANAB process. These results correlate well with those from transmission electron microscopy and X-ray reflectometry. Since ANAB is broadly applicable to material surfaces, the present study indicates its promise for modification of the surfaces of superconducting quantum circuits to achieve longer coherence times.

**Introduction**

In recent years, a major component of microwave loss in state-of-the-art quantum devices has been attributed to native oxides that grow on various materials upon exposure to ambient air [1, 2]. Niobium (Nb) has been widely used in the fabrication of superconducting single-flux quantum circuits [3], and as the material for interconnects, feedlines and coplanar waveguide resonators in multi-qubit quantum computing systems [4]. However, many researchers have noted that Nb forms a complex native oxide composed of a mixture of $Nb_2O_5$, $NbO_2$, $NbO$, and other sub-oxides [5, 6]. Verjauw *et al* reported that $Nb_2O_5$ grows logarithmically over time up to 7 nm, and estimated the loss tangent of $Nb_2O_5$ to be about $1 \times 10^{-2}$ [2]. A 100 nm thick Nb resonator with intact native oxide showed a quality factor (Q) of $1 \times 10^6$, but improved to $7 \times 10^6$ when the native oxide was etched away [2]. Not unexpectedly, they find that the native oxide regrows over time, with high-Q resonators showing microwave loss dominated by TLS defects in $Nb_2O_5$ after 6 hours [2]. The loss mechanism associated with regrown native oxide is important enough that even rapid UHV packaging after sputter cleaning is being explored [7]. Therefore, processes that create stable surfaces that can survive ambient air exposure without degrading Q are desirable. Zheng *et al* studied nitrogen doping of Nb surfaces by plasma treatment, showing a four-fold decrease in loss tangent [8]. They also reported that XPS spectra of nitrogen-passivated Nb surfaces remain stable after aging for 15 days. However, the resonator Q was measured after etching away the plasma-nitrided Si surfaces alongside the Nb resonator. While this clarifies the effect of Nb passivation in their study, it also illustrates the need for a process that could simultaneously passivate both the superconducting surface as well as the Si surface. A strategy compatible with CMOS fabrication facilities is to engineer the formation of thin, thermally-stable, controlled-stoichiometry oxides through Accelerated Neutral Atom Beam (ANAB) technology [9, 10]. This ANAB technique is also useful to variety of materials (e.g., Ti, Cu, Si, sapphire, teflon) for precise sputtering, nano-level surface smoothing, extremely shallow doping, highly selective etching, ultra-thin film deposition, surface passivation, various surface specific molecular transformations, and for a range of other ultra-shallow surface actions [9-11]. Due to the low energy of the neutral atoms, ANAB enables uniform nanoscale depth processing of surfaces that is required for quantum applications.

Using the ANAB process, we developed an innovative approach to engineer the niobium oxide structures at room temperature to form a stable and engineered (2 nm to 6 nm thick) Nb-oxide by changing oxygen fluence. In the present work, we studied the thickness and stoichiometry of the various niobium oxides ($Nb_xO_y$) formed by oxygen doped ANAB treatment of Nb films and its variation with oxygen fluence. To characterize the nature of the Nb oxides, we used synchrotron X-ray photoelectron spectroscopy (XPS) at NSLS-II beamlines at Brookhaven National Laboratory (BNL) with variable photon energy. Our transmission electron microscopy (TEM) results agree well with the XPS depth analysis and XRR results and successfully demonstrate engineering of thickness and stoichiometry of Nb surface oxides. The initial results from ANAB treated Nb samples look promising and underscore its potential to be used for fabricating high-Q resonators for quantum applications.

*Accelerated Neutral Atom Beam (ANAB) technology*

ANAB is a novel process that uses an intense directed beam of neutral gas atoms with average energies ranging from ~5 to ~100 eV/atom [9, 10]. Adiabatic expansion through a specially designed nozzle into an ultra-high vacuum causes the gas to form clusters. The clusters are ionized as shown in Figure 1, and accelerated by an electric field. The neutral atom beam, directed at the wafer surface, is produced by breaking up the accelerated clusters into atoms, while steering ions and charged sub-clusters away.

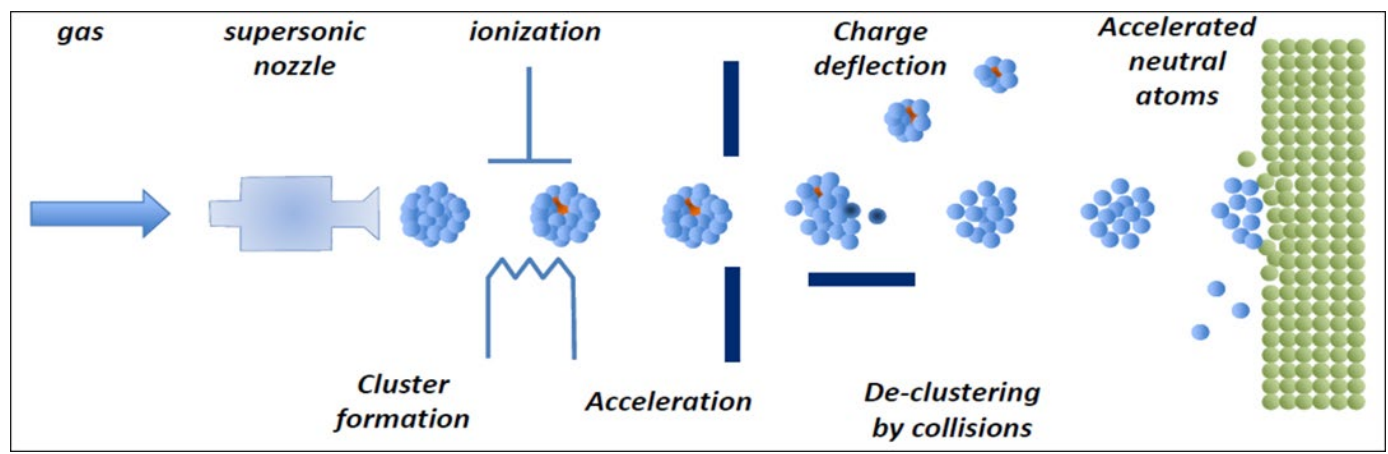

**Figure 1.** *Surface engineering using accelerated neutral atom beams (ANAB).*

The low energy, neutral atoms of ANAB reduce charging concerns, limit sub-surface crystalline damage, and limit surface modification to nanometer scales. Results of 30 kV Ar-based ANAB experiments on Si surfaces have shown that 2.1 nm thick, high-quality amorphous $SiO_2$ can be formed at room temperature with an atomically smooth interface [9, 11]. ANAB has shown promise for metal surfaces such as Cu, where the formation of ANAB Cu oxide prevented any further growth of the Cu oxide for several months [10]. These effects motivated the study of ANAB-modified superconducting metal films with an eye to engineering the surface oxide for use in superconducting quantum circuits with lower decoherence.

**Experimental**

*Sample preparation on 300 mm wafers*

For the preparation of the samples, a $SiO_2$ (50nm) / SiN (20 nm) bilayer was deposited on 300 mm diameter Si (100) wafers. A set of 1 µm deep trenches were formed in a periodic array on the wafer surface, to act as fiducials. As shown in Figure 2(a), such fiducials permit high accuracy navigation to find and characterize the ANAB treated and untreated areas. They were located at the bottom left corner of each 825 $mm^2$ exposure field (occupying ~3 $mm^2$), and created using a sequence of 193 nm optical lithography, reactive ion etching (RIE), and post-RIE wet-cleans. A 3 nm thick TaN film was deposited using atomic layer deposition (ALD), followed (without vacuum break) by sputtering of a Nb film. The wafer was then processed through a Chemical Mechanical Planarization (CMP) tool to obtain a smooth Nb surface using a commercially available soft CMP pad and slurry (used for copper diffusion barrier metal removal). While the Nb thickness was not measured before CMP, post CMP Nb thickness was 370 nm. Since CMP slurries typically contain an oxidizing agent such as hydrogen peroxide, the process creates a surface oxide – in this paper, 'untreated' oxide refers to the oxide left after the CMP process. A schematic cross-section of the as-prepared Nb wafer is shown in Figure 2(b). On each of 3 wafers, three zones (12 mm wide and 46.5 mm high) were treated with ANAB processes at different oxygen fluences (Table I) at the locations shown in Figure 2(a). Each wafer used a gas different flow rate for the ANAB process (low, medium and high – labeled LF, MF and HF).

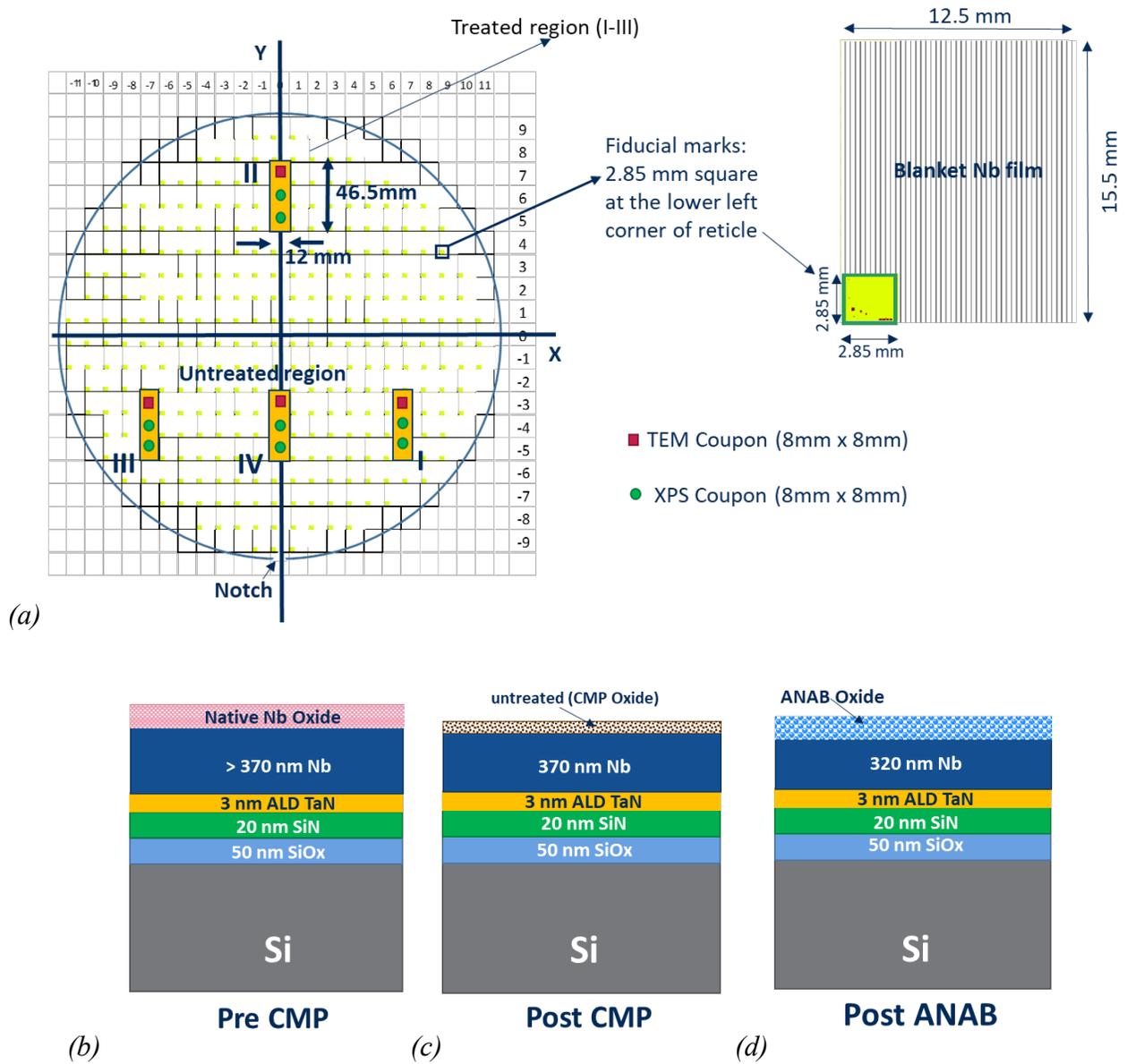

**Figure 2.** *(a) ANAB treated and untreated zones together with fiducial marks on the 300 mm wafers to identify the die (each die size is 12.5 mm × 15.5 mm) locations, (b) schematic cross-section of the as-prepared Nb wafer with native oxide layer, (c) schematic cross-section of the post CMP Nb wafer with CMP oxide layer (d) schematic cross-section of the ANAB treated wafers where Nb oxide formed during CMP is replaced with an engineered and stable oxide. It should be noted that all schematics (b, c, d) are not drawn to scale.*

*Sample characterization*

In the ANAB tool, 50 nm of the top surface was first removed by an Ar-only neutral atom beam to sputter away native oxide of niobium and sufficient Nb metal to eliminate any questions about sputter-mixed O atoms that remain in the Nb. Immediately after that (without any vacuum break), the ANAB gas input was switched to an 90% Ar/10% $O_2$ mixture. ANAB oxygen fluence was separately determined by recording the pressures created by the beam directed at a specially designed collector. The pressure in the collector is measured by a hot cathode ion gauge, with ±10% accuracy, hence limiting fluence determination to ±10%. An ANAB-formed Nb oxide layer was formed in specific regions of the wafer by rastering the beam over the region with conditions designed to expose the Nb to a chosen value of oxygen fluence. The ANAB conditions pertinent to each of the 3 regions on each of the 3 wafers are provided in Table I. Schematic cross-sections (not drawn to scale) are shown in Figures 2 (b-d) corresponding to various stages of the sample preparation process.

Table I - *Specifications of ANAB treated samples (ANAB oxygen fluence has ±10% error)*

| Zone | ANAB treated | Flow condition | ANAB oxygen fluence (atoms.cm$^{-2}$) |
|---|---|---|---|
| I | Yes | Low flow (LF) | $2.34 \times 10^{15}$ |
| | | | $3.9 \times 10^{15}$ |
| | | | $5.46 \times 10^{15}$ |
| II | | Medium flow (MF) | $4.07 \times 10^{15}$ |
| | | | $6.78 \times 10^{15}$ |
| | | | $9.49 \times 10^{15}$ |
| III | | High flow (HF) | $1.14 \times 10^{16}$ |
| | | | $1.89 \times 10^{16}$ |
| | | | $2.65 \times 10^{16}$ |
| IV | No | No ANAB | 0 |

After ANAB exposure, we carried out XRR measurements on full 300 mm wafers at the ANAB treated and untreated regions (yellow boxes in Figure 2(a)). Since glancing incidence of the XRR beam elongates its 'footprint' on the wafer surface, the XRR thickness measurement was made with the beam aligned parallel to the y-axis (Figure 2(a)). After XRR measurements were completed, the wafer was cleaved to get the ANAB treated and untreated coupons of 8 mm × 8 mm size (red squares in Figure 2(a)) for TEM measurements. Each wafer has three ANAB treated regions and untreated regions, resulting in 4 coupons from each wafer for TEM sample preparation. A thick (~50-60 nm) carbon capping layer was used to protect the top surface of Nb oxide film during Focused Ion Beam (FIB) preparation of the TEM lamellae, including a 2-3 nm thick Iridium (Ir) coating beneath the thick carbon capping layer for thicker Nb-oxide samples. Wedge shaped TEM lamellae were prepared by FIB so that the thinnest region of the lamella that was not amorphized could be imaged, to obtain the best TEM image.

For XPS characterization, we similarly cleaved 12 square shaped (8×8 mm$^2$) coupons, along with 12 more coupons as backup samples (green dots) as shown in Figure 2(a).

XPS measurements were made at the HAXPES end-station on the NIST Spectroscopy Soft and Tender (SST) beamline suite at National Sychrotron Light Source II (NSLS-II), using both SST-1 ($h\nu$ < 2000 eV) and SST-2 ($h\nu \geqslant$ 2000 eV) beams. The end-station is equipped with a Scienta R4000 concentric hemispherical analyzer with 200 mm radius and using a 300 μm curved analyzer entrance slit. The analyzer was operated in angular mode [12], but spectra were integrated over the angular axis. The analyzer is mounted perpendicular to the SST-2 beam propagation direction with the central analyzer axis parallel with the SST-2 beam polarization vector. For the SST-1 beam, the angle between beam propagation and analyzer is approximately 94° with an approximately 4° angle between x-ray polarization vector and analyzer axis.

For all measurements, samples were mounted with an 80° takeoff angle. The x-ray incident angle was thus 10° for SST-2 beam and 6° for SST-1 beam. The photon energy was calibrated for each photon energy by measuring the Ag $3d_{5/2}$ core line from an Ag foil mounted with the samples and setting the binding energy to 368.3 eV. Spectra were taken using a point spacing of 0.05 eV and a dwell time of 0.1 seconds.

Energy selection for the SST-1 beam was accomplished using a 1200 l/mm variable line spacing plane grating monochromator with a 25 μm exit slit for an estimated resolving power ($E/\Delta E$) of $10^4$. For SST-2, a double-crystal monochromator is used. The 2000 eV and 6000 eV photon energies were taken using a Si (111) crystal pair with the first and third order beams respectively. Intermediate energies were collected using a Si (220) crystal pair. The overall experimental broadening will depend on photon energy and analyzer pass energy but should be less than 0.35 eV for all conditions. The overall experimental broadening at 2000 eV photon energy was measured as the Gaussian broadening of the Fermi level from a silver foil and found to be approximately 0.2 eV, meaning a resolving power of $10^4$.

**Results and Discussion**

*Cross-sectional transmission electron microscope image analyses*

TEM cross-sections of samples treated by ANAB under HF, MF and LF conditions as well as untreated samples are shown in Figure 3(a), 3(b), 3(c) and 3(d) respectively – it can be seen that an increase in the thickness of ANAB-formed oxide is observed with an increase in oxygen fluence. Figure 3(a-d) are taken 30 days after ANAB treatment. Figure 3(a) can be compared with Figure 3(e) which shows the same process condition but with the TEM image taken 100 days after ANAB treatment – suggesting that the ANAB oxide thickness could be similar across the wafer surface and unchanged over 70 days. Comparison of untreated oxide TEM images in Figure 3(d) and 3(f), on the other hand, illustrates that the untreated oxide thickness varies from point to point, since Figure 3(f) is an image from a different untreated area (imaged 100 days after treatment). Multiple measurements were made in each TEM image to estimate the standard error, which was determined to be ≤12%.

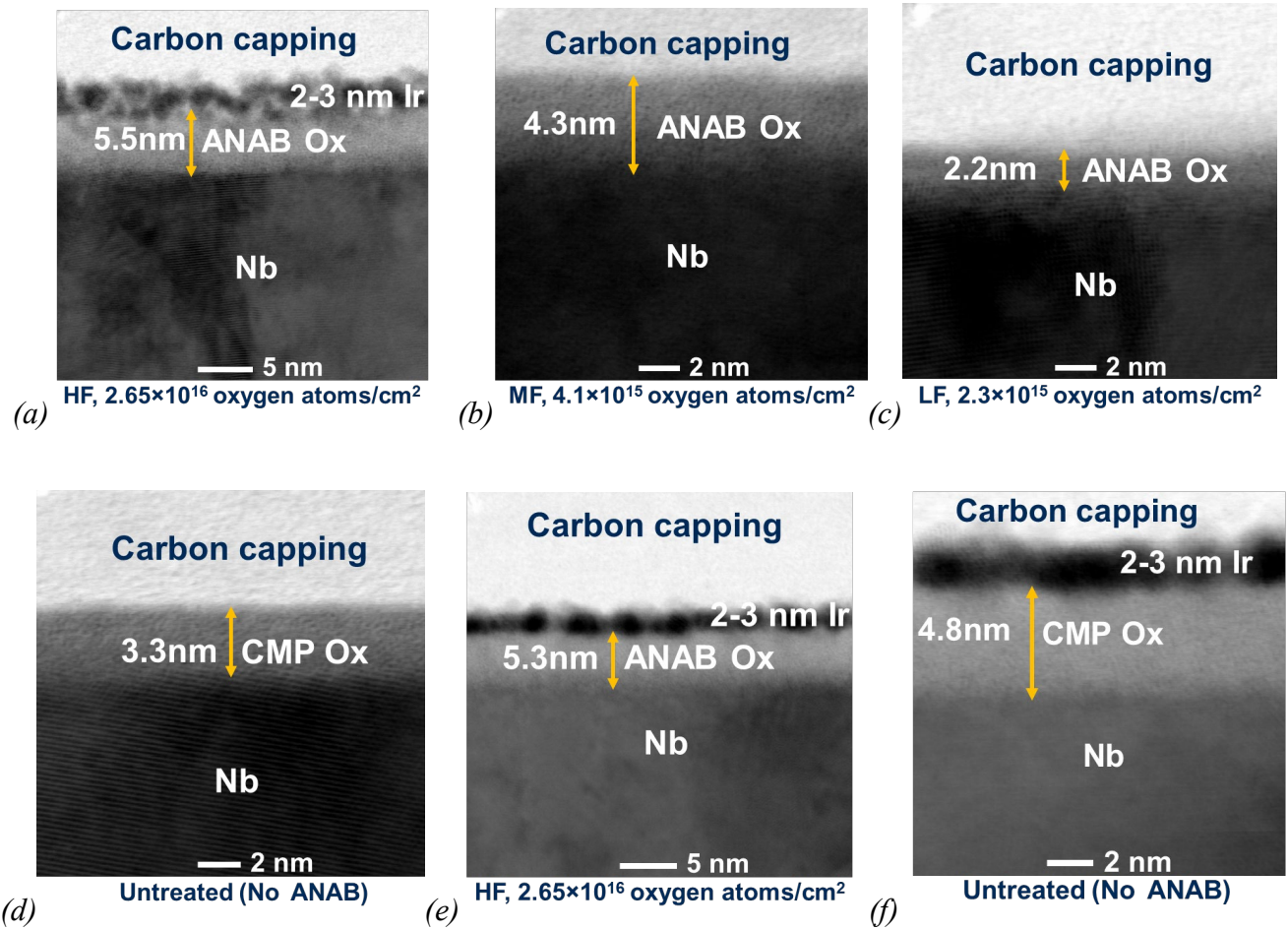

**Figure 3.** *TEM cross-sections (imaged 30 days after ANAB treatment) of (a) HF sample with 2.65×10¹⁶ oxygen atoms/cm² fluence, (b) MF sample with 4.1×10¹⁵ oxygen atoms/cm² fluence, (c) LF sample with 2.3×10¹⁵ oxygen atoms/cm² fluence, (d) untreated (no ANAB) sample, (e) HF sample with 2.65×10¹⁶ oxygen atoms/cm² fluence (imaged 100 days after ANAB treatment) and (f) untreated (no ANAB) sample (imaged after 100 days) from another untreated location.*

*X-ray photoelectron spectroscopy data analysis*

With the ability to vary the incident photon energy, it was possible to identify the relative prevalence of the various niobium oxidation states as a function of depth. All spectra were analyzed using the CasaXPS software package [13]. The narrow-band high-resolution photoemission spectra of interest in the present work focused on Nb-3d core-level region, to identify the chemical shifts of the Nb-3*d* core level when niobium is bonded to oxygen. The following spectrum-fitting procedure was utilized, starting with background subtraction using Shirley-type background with offsets [14]. In the fitting model we considered the binding-energy shifts (ΔBE) of niobium oxides found in the literature and the NIST XPS Database [15]. A well-known linear dependence of the ΔBE of niobium oxides versus niobium valence was considered as a starting point [16, 17]. The core-level peak features were decomposed as doublets with a fixed spin-orbit splitting of 2.75 eV and branching ratio of 3:2 between $3d_{5/2}$ and $3d_{3/2}$. The unoxidized metallic Nb peaks were fitted with the standard CasaXPS asymmetric lineshape, representing a numerical convolution of a Lorentzian with a Gaussian function with asymmetry (LA(0.9,1.8,12), as described in ref. [18]). Other Nb compounds having metallic conductivity ($Nb_2O_{0.4}$, $Nb_2O_{0.6}$, $Nb_2O$, $NbO$, and $Nb_2O_3$) were also fitted with the standard CasaXPS asymmetric lineshape but with different parameters, LA(1.2,5,12). Oxides with semiconducting or dielectric characteristics ($NbO_2$, $Nb_2O_5$) were fitted with a Gaussian/Lorentzian product lineshape GL(50) [19, 20]. Figure 4(a) shows a representative spectrum of the Nb $3d_{3/2}$ and $3d_{5/2}$ core levels, (including fitting to various oxidation states) for the ANAB treated film (HF, oxygen fluence of $2.65 \times 10^{16}$ atoms/cm$^2$) with a photon incident energy of 3500 eV. The XPS spectra are best fit with eight doublet components, revealing that there are seven coexisting oxidation states besides Nb metal: $Nb_2O_{0.4}$ and $Nb_2O_{0.6}$, $Nb_2O$, $NbO$, and $Nb_2O_3$, $NbO_2$ and $Nb_2O_5$, with respective shifts in binding energy of +0.39 eV, +0.60 eV, +1.09 eV, +1.99 eV. +3.14 eV, + 4.12 eV and +5.66 eV compared to Nb metal with a standard error of ±0.05 eV. These BE shift data are in good agreement with the reference [18].

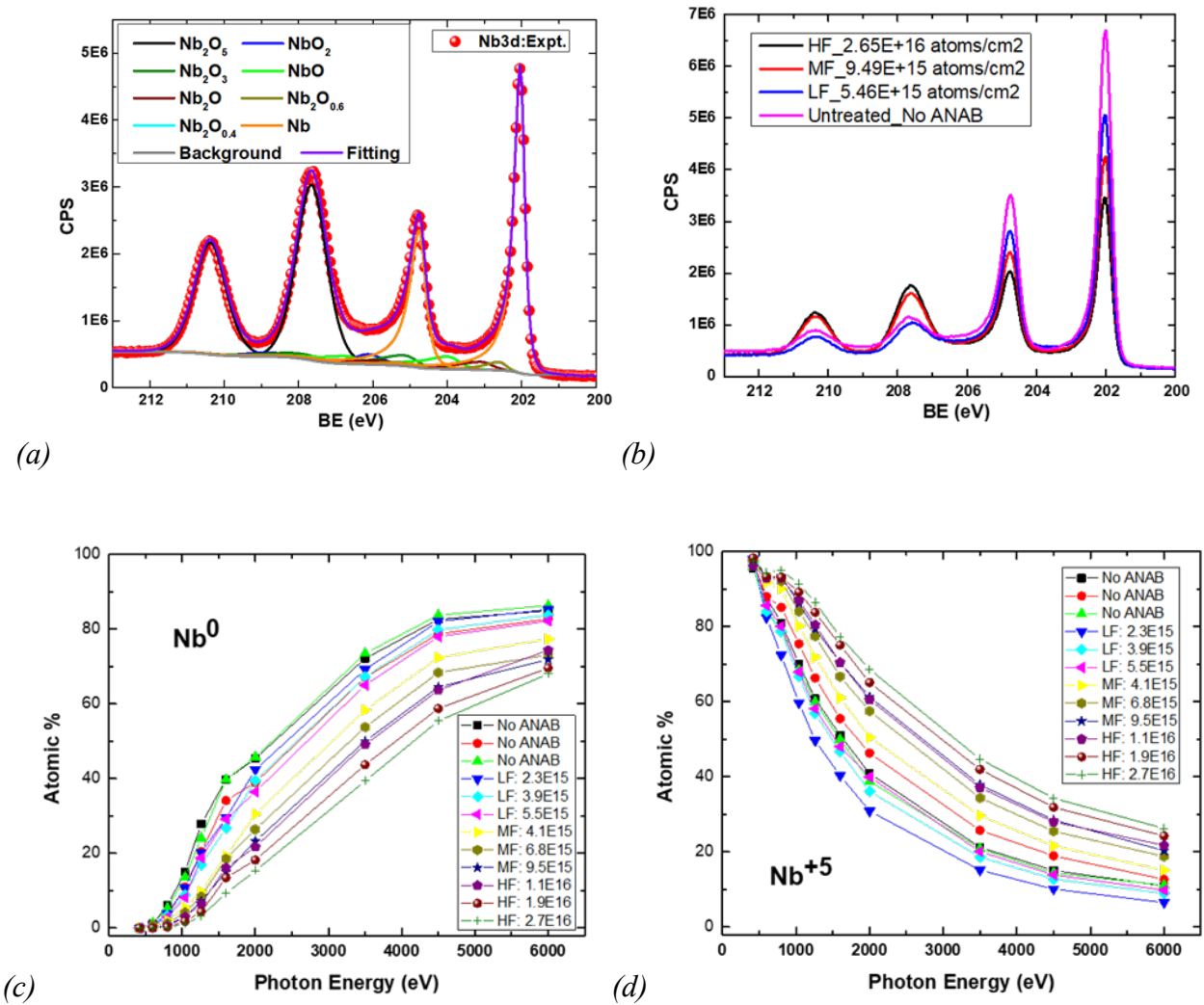

**Figure 4.** *(a) XPS spectrum of the spin-orbit-split Nb doublet ($3d_{3/2}$ and $3d_{5/2}$) core levels and fitting to other doublet material components, measured on the ANAB treated film (HF, oxygen fluence of $2.65 \times 10^{16}$ atoms/cm$^2$) with an incident energy of 3500 eV, (b) spectra measured on all three wafers (HF, MF and LF conditions) for the ANAB oxide film and on no ANAB sample at 4500 eV of photon energy, (c) variation in atomic% of metallic Nb ($Nb^0$) as a function of photon energy and (d) $Nb_2O_5$ ($Nb^{+5}$) respectively as a function of photon energy.*

Figure 4(b) shows the spectra measured on all three wafers (HF, MF and LF conditions) for the ANAB oxide film at 4500 eV, revealing clear variations in the concentrations of $Nb_2O_5$ oxidation states when compared with the untreated sample (Zone IV). Figure 4(c) and 4(d) depict the variation in atomic% of metallic Nb ($Nb^0$) and $Nb_2O_5$ ($Nb^{+5}$) respectively as a function of photon energy. It is observed that metallic Nb ($Nb^0$) atomic% gradually increases and $Nb_2O_5$ ($Nb^{+5}$) atomic% decreases with the increase in photon energy.

*Oxide stoichiometry and thickness model*

The depth profiles of Nb species are simulated using the exponential attenuation model, in which two major kinds of attenuation are involved, i.e., the attenuation of incident photons and the attenuation of generated photoelectrons [21]. Each kind of attenuation has its own attenuation term, and the total attenuation throughout the XPS process is exponentially added by the two terms. For simplicity, the multicomponent thin film is assumed to be homogeneous along the in-plane direction. At the depth of $x$, the attenuation term of incident photon ($T_{ph}$) is the integral over the differential of photon flux across a thin slab of thickness d$s$, as

$$T_{ph}(x) = \int_0^x \frac{\mu(E_{ph})}{\rho_0} \frac{\rho(s)}{\sin\Theta_{ph}} ds, \quad (1)$$

where $E_{ph}$ is the energy of incident photon, $\mu(E_{ph})/\rho_0$ is the tabulated mass attenuation coefficient as a function of $E_{ph}$ in [22], $\rho(s)$ is the total mass density at depth $s$, and $\Theta_{ph}$ is the incident angle. On the other hand, the attenuation term of photoelectron ($T_e$) is determined by the travel distance through the film after photoelectron generation and the effective attenuation length, as

$$T_e(x) = \frac{x}{\lambda_{el} \sin\Theta_{el}}, \quad (2)$$

where $\lambda_{el}$ is the effective attenuation length and $\Theta_{el}$ is the photoelectron collection angle. $\lambda_{el}$ is dependent on the kinetic energy of photoelectron $E_k = E_{ph} - E_b$, in which $E_b$ is the binding energy. For different Nb species, the variance of the binding energies of Nb 3d peaks is only a few eV, which is far less than $E_{ph}$. Thus, $E_b$ is set as a constant value of 207 eV in the model, and $\lambda_{el}$ is assumed to be only dependent on the incident photon energy. The values of $\lambda_{el}$ are converted from the calculated electron inelastic mean free path tabulated in [23]. The total attenuation factor ($T$) is the sum of $T_{ph}$ and $T_e$, as

$$T(x) = T_{ph} + T_e = \int_0^x \frac{\mu(E_{ph})}{\rho_0} \frac{\rho(s)}{\sin\Theta_{ph}} ds + \frac{x}{\lambda_{el} \sin\Theta_{el}}, \tag{3}$$

Within a thin slab of thickness dx at the depth of x, the contribution to the photoelectron signal by the nth Nb species ($dA_n$) is

$$dA_n = \gamma\eta\rho_n V_n(x)\exp(-T(x)), \tag{4}$$

where $\gamma$ is the photoelectron collection efficiency, $\eta$ is the photoelectron yield, $\rho_n$ is the mass density of the nth Nb species, and $V_n(x)$ is the effective volume of the nth Nb species. The $\gamma$ and $\eta$ are both assumed to be independent of species and photon energy. To calculate $V_n(x)$, we define a volume fraction profile $F_n(x)$ for each depth x. The sum of $F_n(x)$ is constrained to be unity ($\sum_{n=1}^N F_n(x) = 1$), and the $F_n(x)$ for the Nth Nb species in material bulk follows the limiting behavior of $\lim_{x\to\infty} F_N(x) = 1$. Based on the volume fraction, $V_n(x)$ is given by

$$V_n(x) = F_n(x)Sdx, \tag{5}$$

where $S$ is the film area. Thus, $dA_n$ can be expressed as

$$dA_n = \gamma\eta S\rho_n F_n(x)\exp\left(-\int_0^x \frac{\mu(E_{ph})}{\rho_0} \frac{\rho(s)}{\sin\Theta_{ph}} ds - \frac{x}{\lambda_{el} \sin\Theta_{el}}\right) dx, \tag{6}$$

The total contribution of the $n^{th}$ Nb species ($A_n$) is produced by the integration of $dA_n$ over x, as

$$A_n = \gamma\eta S\rho_n \int_0^\infty F_n(x)\exp\left(-\int_0^x \frac{\mu(E_{ph})}{\rho_0} \frac{\rho(s)}{\sin\Theta_{ph}} ds - \frac{x}{\lambda_{el} \sin\Theta_{el}}\right) dx. \tag{7}$$

Finally, the $n^{th}$ Nb species contributes a normalized weight to the final spectrum as

$$W_n = \frac{A_n}{\sum_{n=1}^N A_n} = \frac{\rho_n \int_0^\infty F_n(x)\exp\left(-\int_0^x \frac{\mu(E_{ph})}{\rho_0} \frac{\rho(s)}{\sin\Theta_{ph}} ds - \frac{x}{\lambda_{el} \sin\Theta_{el}}\right) dx}{\sum_{n=1}^N \rho_n \int_0^\infty F_n(x)\exp\left(-\int_0^x \frac{\mu(E_{ph})}{\rho_0} \frac{\rho(s)}{\sin\Theta_{ph}} ds - \frac{x}{\lambda_{el} \sin\Theta_{el}}\right) dx}, \tag{8}$$

in which the independent variables are cancelled out. The distribution of $F_n(x)$ across the film, i.e., the depth profile, is obtained by fitting $W_n$ to the measured fractions of XPS peaks.

To reduce the number of fitting parameters, a basis set is needed for $F_n(x)$. Herein, a smooth basis set containing monotonic profiles for the 1st and $N^{th}$ species and unimodal profile for the intermediate species is generated based on a set of $N$ precursor functions

$$f_n(x) = \begin{cases} H(x) & (n=1) \\ \left[1 + \exp\left(\frac{d_n-x}{w_n}\right)\right]^{-1} & (n>1) \end{cases}, \quad (9)$$

where, $H(x)$ denotes the unit step function, and $d_n$ and $w_n$ are precursor parameters. It is noted $d_n$ is constrained to be not descending, i.e., $d_1 \leq d_2 \leq d_3 \cdots \leq d_N$. The basis set of volume fraction functions are thus generated in the form of the product of precursors, as

$$F_n(x) = \begin{cases} (1 - f_{n+1}(x))\prod_{i=1}^{n} f_i(x) & (n<N) \\ \prod_{i=1}^{N} f_i(x) & (n=N) \end{cases}. \quad (10)$$

As such, the basis set is self-constrained that the sum of $F_n(x)$ is always unity ($\sum_{n=1}^{N} F_n(x) \equiv 1$). The number of fitting parameters is then reduced to $2N-2$, including $N-1$ $d_n$'s and $N-1$ $w_n$'s.

Figures 5 (a-c) show the component fraction of $Nb_2O_5$, other Nb oxides ($Nb_2O_{0.4}$, $Nb_2O_{0.6}$, $Nb_2O$, NbO, $Nb_2O_3$, $NbO_2$), and metallic Nb ($Nb^0$) at different photon energies for no ANAB, LF and HF conditions respectively. The resulting depth profiles of the relative concentrations of $Nb_2O_5$, the other Nb oxides, and metallic Nb ($Nb^0$) states are shown in Figure 5 (d-f). From Figure 5 (a-f), we observe that all three conditions (no ANAB, LF and HF) have a surface oxide layer with $Nb_2O_5$ as the main constituent, followed by a transition layer (composed primarily of other Nb oxides) whose thickness, depth distribution, and composition varies between the different conditions. The oxide/metal interface has roughly the same sharpness for all conditions tested. The HF condition (2.65 × $10^{16}$ atoms/cm$^2$ oxygen fluence) creates the thickest $Nb_2O_5$ layer, with an intermediate region dominated by other niobium oxides that extends from ~ 3 nm to ~ 5 nm below the surface. The LF condition (5.46 × $10^{15}$ atoms/cm$^2$ oxygen fluence) shows the thinnest $Nb_2O_5$ layer that is formed by ANAB under the conditions studied, similar to the $Nb_2O_5$ thickness seen with CMP-formed oxide. However, in contrast to CMP-formed oxide, the transition layer (with other oxides of Nb) is sharply defined, and extends only between ~ 2 nm and ~ 3 nm below the surface.

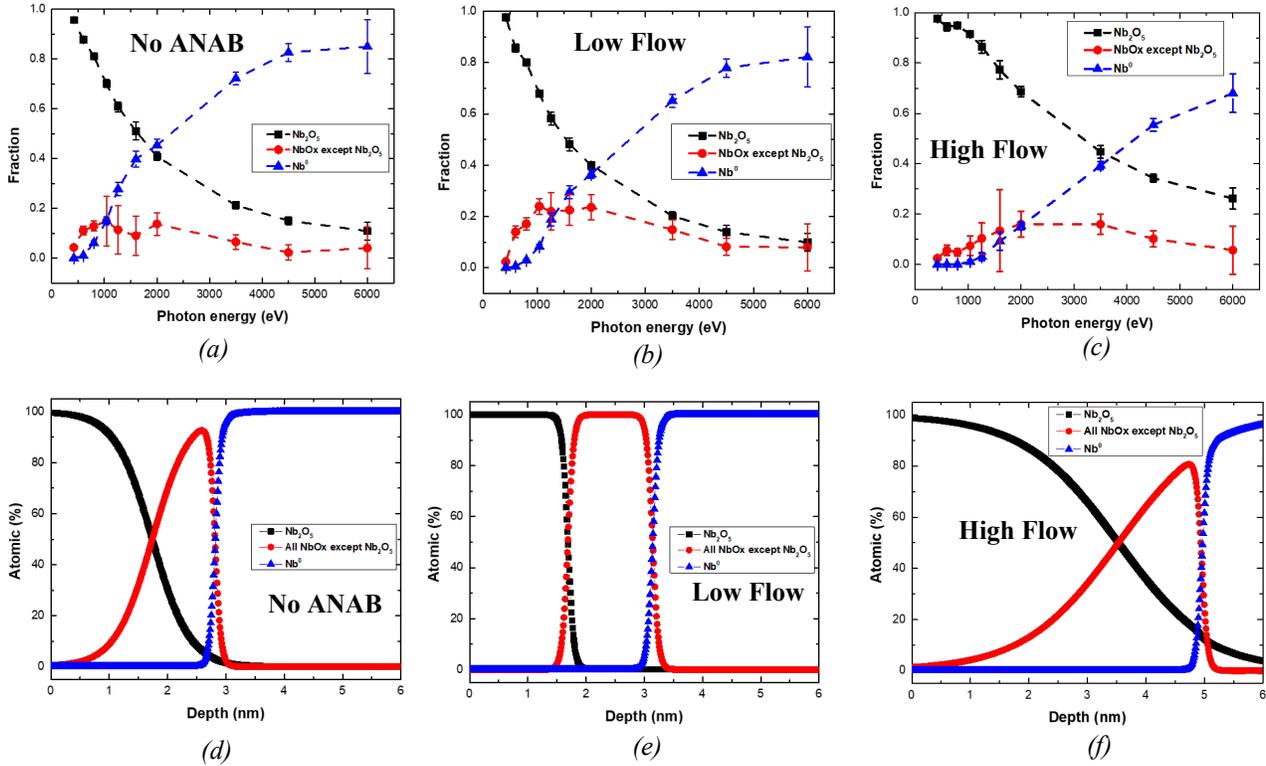

**Figure 5.** *Component fractions of $Nb_2O_5$, other Nb oxides, and metallic Nb ($Nb^0$) at different photon energies for (a) untreated (No ANAB), (b) LF (5.46 × $10^{15}$ atoms/cm$^2$ oxygen fluence) and (c) HF (2.65 × $10^{16}$ atoms/cm$^2$ oxygen fluence). Atomic % vs depth profiles of $Nb_2O_5$, other Nb oxides, and metallic Nb ($Nb^0$) for (d) untreated (No ANAB), (e) LF (5.46 × $10^{15}$ atoms/cm$^2$ oxygen fluence) and (c) HF (2.65 × $10^{16}$ atoms/cm$^2$ oxygen fluence).*

Figure 6 provides a comparison of the total Nb oxide thickness determined by TEM and XPS along with $Nb_2O_5$ thickness from XPS at different oxygen fluences. The plot also indicates how the oxide thickness trends as function of ANAB oxygen fluence at different flow conditions, ranging from a minimum of 3 nm to a maximum of 5 nm for the process conditions studied. For both MF and HF conditions, the oxide thicknesses from TEM and XPS are in good agreement with each other. For the LF case, XPS modeling provides thickness that underestimate TEM values (by < 1 nm). In the untreated case, both TEM and XPS show the possibility of significant variation (over 1 nm). The difference between TEM and XPS could also arise from spatial variation in the nature of the oxide, since measurements at the same point are not feasible, and TEM from different locations differ in thickness (as was noted earlier). It is of particular interest to note that all the ANAB-formed oxides (as well as the CMP-formed oxide) are substantially thinner than the 5 – 7 nm thick native oxide [2] that forms on sputtered niobium films upon exposure to air ambient over the course of several days.

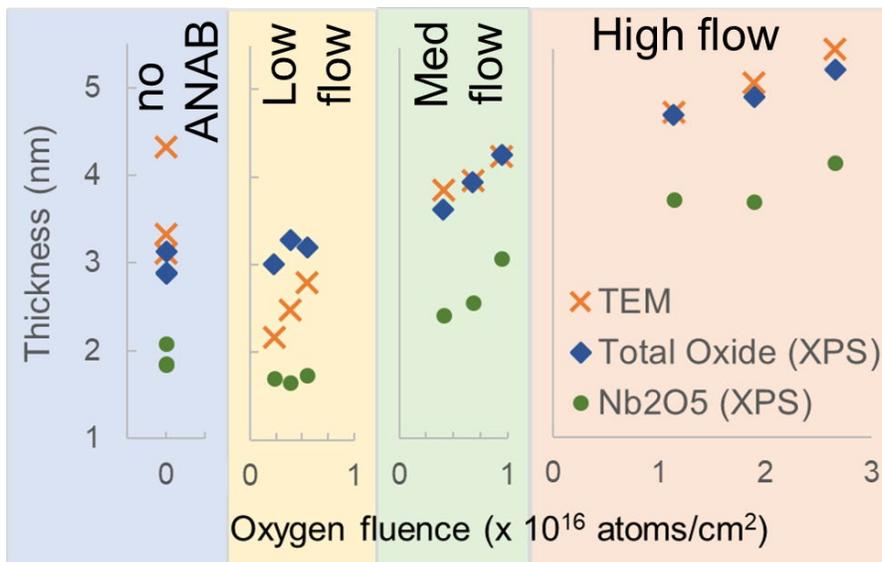

**Figure 6.** *Nb total oxide thickness comparison from TEM and XPS along with $Nb_2O_5$ thickness from XPS. The standard error in TEM thickness is ≤12%.*

*X-ray reflectometry data verification*

Figure 7 presents XRR data for ANAB-treated and untreated cases along with the results of modeling (and constrained fitting). Due to the complex nature of surface oxides of Nb, it was not possible to fit the XRR data with a single oxide layer on Nb. As described above, the sharp interfaces of the transition layer ('other Nb oxides') for LF ANAB case permits modeling with a simpler layer stack. Therefore, we used a 2-layer oxide ($Nb_2O_5$ as the top layer and $NbO_x$ as the layer underneath) on a Nb film of a fixed thickness (320 nm) with Si as the substrate for LF conditions. The thicknesses obtained from the depth analysis were taken as the initial values for XRR fitting using an algorithm provided by Bruker. The same model was used for the untreated Nb case with the thickness of Nb fixed at 370 nm. Resultant fits are shown in Figure 7 along with the raw data. It can be seen that the untreated case fits well with the model as the untreated Nb oxide consists of a mixture of suboxides. As the boundary between $Nb_2O_5$ and $NbO_x$ is not sharply defined for the MF and HF cases, a 3-layer oxide model was used ($Nb_2O_5$ as the top layer, and two oxide layers, $NbO_x$ A and $NbO_x$ B, were used to define the transition region) on Nb. XRR fits were improved when the $NbO_x$ layer thickness and density were allowed to float. The initial values of $NbO_x$ and $Nb_2O_5$ oxide thicknesses were taken from the thickness values obtained from XPS. XRR & XPS model results are consistent with each other. The total thickness of niobium oxide is within ~1.5 nm of the XPS determined value.

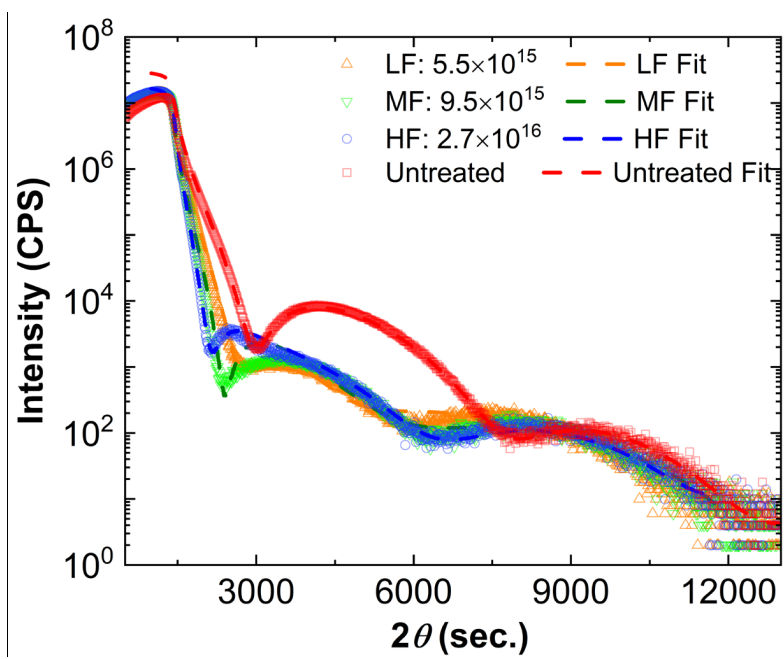

**Figure 7.** *XRR data for ANAB treated and untreated (No ANAB) samples with their fitting models.*

**Conclusions**

In the present work, we have successfully demonstrated the creation of an engineered and stable oxide of Nb, using room-temperature ANAB process in 300 mm wafer scale. Using synchrotron XPS with variable photon energy coupled with modeling, we quantified the thickness and stoichiometry of the various niobium oxides ($Nb_xO_y$) formed by oxygen-doped ANAB on Nb surfaces, and its dependence on oxygen fluence. The best fits to the acquired XPS spectra were obtained by invoking seven oxides in addition to the underlying Nb metal ($Nb_2O_{0.4}$ and $Nb_2O_{0.6}$, $Nb_2O$, $NbO$, and $Nb_2O_3$, $NbO_2$ and $Nb_2O_5$). For HF and MF conditions, ANAB oxide thickness obtained from XPS and TEM showed good agreement (< 6% difference). However, the LF condition showed differences of the range 0.2 to 1 nm. The total thickness of niobium oxide obtained from XRR is within ~1.5 nm of the XPS determined value. The thickness of oxides engineered by ANAB are thinner than the native oxide that is formed on sputtered niobium, and which is regenerated to the same thickness over the course of several days after etching it away. The TEM images obtained after 30 days and 100 days of ANAB treatment, suggest that ANAB oxide thickness is stable over time. The lower thickness and stability over time suggest that the native oxide can be successfully replaced with ANAB-formed oxides. These initial results on Nb-ANAB treated samples look promising for further trial of novel room temperature ANAB treatment on niobium-based devices for quantum applications.

**Acknowledgements**

This material is based upon work primarily supported by the U.S. Department of Energy, Office of Science, National Quantum Information Science Research Centers, Co-design Center for Quantum Advantage ($C^2QA$) under contract number DE-SC0012704, including through sub-contract #390040. This research used resources of the Spectroscopy Soft and Tender Beamlines (SST-1 and SST-2) of the National Synchrotron Light Source II and computational resources of the Center for Functional Nanomaterials (CFN), U.S. Department of Energy Office of Science Facilities at Brookhaven National Laboratory under contract no. DE-SC0012704.